\documentclass[twocolumn,showpacs,aps]{revtex4}

\tighten
\input{tcilatex}

\begin{document}

\title{Local quantum critical behavior in magnetic Kondo impurity models
with partially screened moment}
\author{Guang-Ming Zhang$^{1,3}$ and Lu Yu$^{2,3}$}
\affiliation{$^{1}$Department of Physics, Tsinghua University, Beijing 100084, China;\\
$^{2}$ Institute of Theoretical Physics and Interdisciplinary Center of
Theoretical Studies, CAS, Beijing 100080, China;\\
$^{3}$Center for Advanced Study, Tsinghua University, Beijing 100084, China.}

\begin{abstract}
In the strong coupling limit, an effective resonant level model is derived
for the spin-1 underscreened Kondo impurity model. A local quantum critical
behavior is induced by the formation of a bound state with partially
screened magnetic moment, displaying a residual $Z_2$ symmetry, leading to a 
$\delta $-resonance at the Fermi level in the impurity spectral function. As
a consequence, a logarithmic singularity appears in the real part of the
impurity dynamic spin susceptibility as a function of $\max (\omega ,T)$,
with $\omega $ as frequency, $T$ temperature. A small magnetic field breaks
this $Z_2$ symmetry and suppresses the singularity. We also discuss the
possible manifestation of such a quantum critical behavior in the spin-1/2
two Kondo impurity model and the single quantum dot system with even
electron occupation.
\end{abstract}

\pacs{72.15.Qm, 75.20.Hr, 75.30.Mb}
\maketitle

\section{Introduction}

The Kondo model was initially introduced to describe the behavior of diluted
magnetic impurities in metals, and has been thoroughly studied in various
aspects. Localized quantum impurities also serve as building blocks of the
heavy fermion materials, and there is still a strong interest in studying
Kondo model on the lattice, particularly in view of the breakdown of Fermi
liquid behavior due to incomplete Kondo screening close to a quantum
critical point \cite{si}, as discovered in a large class of f-electron
metals such as CeCu$_{6-x}$Au$_{x}$\cite{schroder} and YbRh$_{2}$Si$_{2}$%
\cite{gegenwart}.

Recently, special attention was paid to the study of the underscreened Kondo
problem which showed some evidence of deviation from the standard Fermi
liquid behavior \cite{coleman-2003}. This result was interpreted as due to
anomalous scattering of conduction electrons on a bound state with remaining
unscreened magnetic moment. Although the large-N approaches using both
spin-boson and spin-fermion representations were developed \cite%
{paul-coleman,florens}, there are still some discrepancies between the
obtained results. The more rigorous treatments based on the Bethe ansatz and
numerical renormalization group (NRG) calculations lead to a ``singular''\
Fermi liquid behavior \cite{mehta,hewson}, namely a Fermi liquid fixed point
with singular irrelevant corrections. However, these insightful works \cite%
{mehta} mainly focused on the spinon phase shift and density of states,
yielding \textit{limited} results. In order to investigate the local dynamic
singularity of the partially screened moment, we would like to propose a
simple approach to grasp the essential physics of the strong coupling fixed
point, revealing the origin of the singularity, and to explore the possible
manifestation of such a quantum critical behavior in the spin-1/2 two Kondo
impurity model and quantum dots with even electron occupation.

The paper is organized as follows. In Sec. II, we present our theory for the
spin-1 underscreened Kondo impurity model. With a pseudo-fermion
representation for the spin-1 operator, an effective resonant level model is
derived in the strong coupling mean field (MF) approximation, and the static
and dynamical properties of the effective model are given in detail. In Sec.
III, a similar approach is proposed to treat the spin-1/2 two Kondo impurity
model. From the resulting effective resonant level model, we identify that
the two Kondo impurity model shares the same singular behavior as the spin-1
underscreened Kondo impurity model for the low-energy excitations. In Sec.
IV, we also explore the possible realization of the singular underscreened
Kondo behavior in a single quantum dot system with even number of electrons.
The conclusion of the paper is given in Sec.V.

\section{Exploration of the spin-1 Kondo impurity model}

The isotropic single Kondo impurity model is usually defined as:%
\begin{eqnarray}
H &=&\sum_{\mathbf{k},\sigma }\epsilon _{\mathbf{k}}C_{\mathbf{k},\sigma
}^{\dagger }C_{\mathbf{k},\sigma }+\frac{J}{2}\left( S^{+}C_{\downarrow
}^{\dagger }C_{\uparrow }+S^{-}C_{\uparrow }^{\dagger }C_{\downarrow }\right)
\nonumber \\
&&+\frac{J}{2}S^{z}\left( C_{\uparrow }^{\dagger }C_{\uparrow
}-C_{\downarrow }^{\dagger }C_{\downarrow }\right) ,
\end{eqnarray}%
where $J>0$ corresponds to an antiferromagnetic coupling. When the impurity
spin $S>1/2$, the strong coupling limit corresponds to formation of a bound
state composed of the conduction electron and the magnetic moment \cite%
{nozieres}. Such a bound state would have a residual degeneracy associated
with the remanent magnetic moment. It is believed that such a partially
screened \textit{quantum} moment should induce a local quantum critical
behavior close to zero temperature \cite{coleman-2003}.

\subsection{Pseudo-fermion representation for spin-1 operator}

As the simplest case, we consider a spin-1 magnetic impurity. Introducing a
pseudo-fermion representation: 
\begin{eqnarray}
S^{+} &=&\sqrt{2}\left( d_{0}^{\dagger }d_{-1}+d_{1}^{\dagger }d_{0}\right) ,
\nonumber \\
S^{-} &=&\sqrt{2}\left( d_{-1}^{\dagger }d_{0}+d_{0}^{\dagger }d_{1}\right) ,
\nonumber \\
S^{z} &=&\left( d_{1}^{\dagger }d_{1}-d_{-1}^{\dagger }d_{-1}\right) ,
\end{eqnarray}%
where $d_{1}$, $d_{0}$, $d_{-1}$ correspond to the three components of $%
S^{z}=1,0,-1$, respectively. To fix the quantum spin magnitude of $S^{2}=2$,
we have to impose the constraint%
\begin{equation}
d_{1}^{\dagger }d_{1}+d_{0}^{\dagger }d_{0}+d_{-1}^{\dagger }d_{-1}=1.
\end{equation}
Since the commutation relations are obeyed $\left[ S^{+},S^{-}\right]
=2S^{z} $ and $\left[ S^{z},S^{\pm }\right] =\pm S^{\pm }$ forming an SU(2)
Lie algebra, this is a \textit{faithful} representation of the quantum
spin-1 operator. Then the model Hamiltonian can be expressed as: 
\begin{widetext}
\begin{equation}
H=\sum_{\mathbf{k},\sigma }\epsilon
_{\mathbf{k}}C_{\mathbf{k},\sigma }^{\dagger }C_{\mathbf{k},\sigma
}-\frac{J}{2}\left( \sqrt{2}d_{1}^{\dagger
}C_{\uparrow }+d_{0}^{\dagger }C_{\downarrow }\right) \left( \sqrt{2}%
C_{\uparrow }^{\dagger }d_{1}+C_{\downarrow }^{\dagger
}d_{0}\right)-\frac{J}{2}\left( \sqrt{2}d_{-1}^{\dagger
}C_{\downarrow }+d_{0}^{\dagger }C_{\uparrow }\right) \left(
\sqrt{2}C_{\downarrow }^{\dagger }d_{-1}+C_{\uparrow }^{\dagger
}d_{0}\right) ,
\end{equation}
\end{widetext}where an irrelevant potential term has been neglected. We
emphasize that employing this pseudo-fermion representation for the spin-1
moment is \textit{instrumental} to explicitly reveal the residual symmetry
of the bound state formed in the strong coupling limit.

\subsection{Effective resonant level model and impurity spectral function}

In analogy with the functional integral approach for the spin-1/2 Kondo
impurity model \cite{read}, a saddle point approximation can be adopted by
defining an effective MF like variable:%
\begin{equation}
\langle \sqrt{2}d_{1}^{\dagger }C_{\uparrow }+d_{0}^{\dagger }C_{\downarrow
}\rangle =\langle \sqrt{2}d_{-1}^{\dagger }C_{\downarrow }+d_{0}^{\dagger
}C_{\uparrow }\rangle \equiv -\sqrt{2}V,
\end{equation}
where we assume the invariance of the low-energy excitations under the
transformations $d_{1}\Leftrightarrow d_{-1}$ and $C_{\uparrow
}\Leftrightarrow C_{\downarrow }$. An effective resonant level model is then
derived: 
\begin{eqnarray}
H_{eff} &=&\sum_{\mathbf{k},\sigma }\epsilon _{\mathbf{k}}C_{\mathbf{k}%
,\sigma }^{\dagger }C_{\mathbf{k},\sigma }+\frac{JV}{\sqrt{2N}}\sum_{\mathbf{%
k,\sigma }}\left( C_{\mathbf{k},\sigma }^{\dagger }d_{0}+h.c.\right) 
\nonumber \\
&&+\frac{JV}{\sqrt{N}}\sum_{\mathbf{k}}\left( C_{\mathbf{k},\uparrow
}^{\dagger }d_{1}+C_{\mathbf{k},\downarrow }^{\dagger }d_{-1}+h.c.\right) 
\nonumber \\
&&+\lambda \sum_{\alpha }d_{\alpha }^{\dagger }d_{\alpha }+2JV^{2}-\lambda ,
\end{eqnarray}
where the constraint has been implemented by a Lagrange multiplier $\lambda $%
. In the effective resonant level model, $d_{1}^{\dagger }|0\rangle $
hybridizes with the spin-up electrons, $d_{-1}^{\dagger }|0\rangle $ with
the spin-down electrons, while $d_{0}^{\dagger }|0\rangle $ hybridizes with
both spin-up and -down electrons. Then the effective resonant level displays
a Z$_{2}$ symmetry describing the invariance of the model under the
transformations $d_{1}\Leftrightarrow d_{-1}$ and $C_{\uparrow
}\Leftrightarrow C_{\downarrow }$, i.e., an Ising symmetry. This property
reflects the residual symmetry of the bound state in the strong coupling
limit and the \textit{quantum} nature of the partially screened moment.

Introducing a Nambu spinor to denote the impurity triplet state $\phi
^{\dagger }=\left( d_{1}^{\dagger },d_{0}^{\dagger },d_{-1}^{\dagger
}\right) $, we can derive the impurity retarded Green's function (GF) matrix
as%
\[
\widehat{G}_{d}(\omega )=\left[ 
\begin{array}{ccc}
\omega -\lambda +i\Gamma & i\Gamma /\sqrt{2} & 0 \\ 
i\Gamma /\sqrt{2} & \omega -\lambda +i\Gamma & i\Gamma /\sqrt{2} \\ 
0 & i\Gamma /\sqrt{2} & \omega -\lambda +i\Gamma%
\end{array}%
\right] ^{-1}, 
\]%
where $\Gamma =\pi \rho _{f}(JV)^{2}$ is a hybridization integral between
the magnetic moment and the conduction electrons. The diagonal retarded GF
is then derived,%
\begin{eqnarray}
G_{1}(\omega ) &=&\frac{\left( \omega -\lambda +i\Gamma \right) \left(
\omega -\lambda +i\Gamma \right) +\Gamma /2}{\left( \omega -\lambda \right)
\left( \omega -\lambda +i\Gamma \right) \left( \omega -\lambda +i2\Gamma
\right) },  \nonumber \\
G_{0}(\omega ) &=&\frac{\left( \omega -\lambda +i\Gamma \right) }{\left(
\omega -\lambda \right) \left( \omega -\lambda +i2\Gamma \right) }, 
\nonumber \\
G_{-1}(\omega ) &=&G_{1}(\omega ),
\end{eqnarray}%
and non-zero off-diagonal retarded GFs are all equal to 
\begin{equation}
G_{1,0}(\omega )=\frac{-i\Gamma /\sqrt{2}}{\left( \omega -\lambda \right)
\left( \omega -\lambda +i2\Gamma \right) }.
\end{equation}%
By summing up the imaginary parts of the diagonal GFs, the impurity spectral
function can thus be obtained%
\[
A_{d}(\omega )=\delta \left( \omega -\lambda \right) +\frac{1}{\pi }\left[ 
\frac{\Gamma }{\left( \omega -\lambda \right) ^{2}+\Gamma ^{2}}+\frac{%
2\Gamma }{\left( \omega -\lambda \right) ^{2}+4\Gamma ^{2}}\right] , 
\]%
where an essential feature of the effective model has been displayed: two
Lorentzian resonances are exhibited at $\omega =\lambda $ with half widths $%
\Gamma $ and $2\Gamma $, respectively; what is more, a sharp $\delta $%
-resonance appears at $\omega =\lambda $. Unlike the two-channel spin-1/2
single Kondo impurity model, where partial impurity degrees of freedom
decouple from the conduction electrons leading to a zero fermionic mode \cite%
{ek}, all three components of the moment here are coupled to the conduction
electrons, and a sharp $\delta $-resonance is induced by the remanent Z$_{2}$
Ising symmetry of the bound state. Using the spectral function, the impurity
contribution to the free energy can be evaluated%
\begin{eqnarray}
\delta F &=&-\frac{1}{\pi }\int_{-D}^{D}d\omega n_{f}(\omega )\left[ \frac{%
\pi }{2}\mathrm{sgn}(\omega -\lambda )+\tan ^{-1}\left( \frac{\Gamma }{%
\lambda -\omega }\right) \right.  \nonumber \\
&&+\left. \tan ^{-1}\left( \frac{2\Gamma }{\lambda -\omega }\right) \right]
+2JV^{2}-\lambda ,
\end{eqnarray}%
where $n_{f}(\omega )=(e^{\beta \omega }+1)^{-1}$ is the Fermi distribution
function. At $T=0$, the saddle point solution leads to $\lambda \approx 0$,
indicating that the sharp $\delta $- resonance and two Lorentzian resonances
lie at the Fermi level with $\Gamma \approx D\exp \left( -\frac{2}{3\rho
_{f}J}\right) $, where $D$ is the half conduction electron bandwidth.

The contribution to the impurity entropy can be calculated using the
relation $S=-\left( {\partial F}/{\partial T}\right) $. At $T=0$, we find
that $S_{\mathrm{imp}}=-k_{B}\ln 2$, implying that the magnetic moment has
been partially screened by the conduction electrons. The total impurity
entropy down to $T=0$ will be $\Delta S_{\mathrm{imp}}=k_{B}\ln (3/2)$,
being consistent with the exact result \cite{mattis}. However, the impurity
contribution to the specific heat is still linear in temperature, in spite
of the $\delta $-resonance appearing in the impurity spectral function at
the Fermi level.

\subsection{Dynamic spin susceptibilities and scattering T-matrices}

To fully reveal the singularity in the effective resonant level model, we
calculate the dynamic impurity spin correlation functions at finite
temperatures. After some algebra, the leading order impurity spin dynamic
spectral function is obtained%
\begin{eqnarray}
&&\ \mathtt{Im}\chi _{d}^{zz}(\omega ,T)  \nonumber \\
&=&-\frac{\Gamma }{4}\left[ \frac{1}{\omega ^{2}+\Gamma ^{2}}+\frac{1}{%
\omega ^{2}+4\Gamma ^{2}}\right] \tanh \left( \frac{\omega }{2k_{B}T}\right),
\end{eqnarray}
and the corresponding real part is thus estimated as $\mathtt{Re}\chi
_{d}^{zz}(\omega ,T)\propto \ln ({\mathrm{\max }\left( \omega ,T\right) }/{%
\Gamma })$. The transverse impurity spin dynamic susceptibility $\chi
_{d}^{-+}(\omega ,T)$ shows a similar behavior. Such a logarithmic
divergence of the impurity spin susceptibility is reminiscent of the
singularity in the overscreened two-channel single Kondo impurity model \cite%
{ek}. In some sense the two-channel overscreened and spin-1 underscreened
cases are somehow ``dual'' to each other: in the former case the singularity
is induced by the ``extra'' screening channel of conduction electrons, while
in the latter case it is due to the ``extra spin channel'' of the impurity %
\cite{florens}.

Now we derive the scattering T-matrix of the conduction electrons. By
calculating $G_{\sigma ,\sigma ^{\prime }}(\mathbf{k,k}^{\prime },i\omega
_{n})$ through their equations of motion, the retarded T-matrices are
derived as%
\begin{eqnarray}
T_{\sigma ,\sigma }(\omega ) &=&V^{2}\left[ G_{1}(\omega )+\sqrt{2}%
G_{1,0}(\omega )+\frac{1}{2}G_{0}(\omega )\right] ,  \nonumber \\
T_{\sigma ,-\sigma }(\omega ) &=&V^{2}\left[ \sqrt{2}G_{1,0}(\omega )+\frac{1%
}{2}G_{0}(\omega )\right] ,
\end{eqnarray}%
where the off-diagonal impurity GFs are involved. The scattering cross
sections of the conduction electrons off the bound state are proportional to%
\begin{eqnarray}
\mathtt{Im}T_{\sigma ,\sigma }(\omega ) &=&\frac{-1}{2\pi \rho _{f}}\left[ 
\frac{\Gamma ^{2}}{\omega ^{2}+\Gamma ^{2}}+\frac{4\Gamma ^{2}}{\omega
^{2}+4\Gamma ^{2}}\right] ,  \nonumber \\
\mathtt{Im}T_{\sigma ,-\sigma }(\omega ) &=&\frac{-1}{2\pi \rho _{f}}\left[ -%
\frac{\Gamma }{2}\delta \left( \omega \right) +\frac{3\Gamma ^{2}}{\omega
^{2}+4\Gamma ^{2}}\right] .
\end{eqnarray}%
For the non-spin-flip scatterings, two Lorentzian resonances are obtained
and the $\delta $-resonance is exactly cancelled, while the $\delta $%
-resonance appears in the spin-flip scattering processes where the residual
magnetic moment of the bound state has to spin-flip as well. Thus, the
remanent moment with Z$_{2}$ symmetry in the underscreened case gives rise
to a singular behavior in both local impurity and conduction electrons
properties.

Compared with the Bethe ansatz and NRG results \cite{mehta}, our resonant
level model serves as an effective model at the singular Fermi liquid fixed
point. Given the singular impurity spin dynamic spectral function, the
retarded self-energy of conduction electrons due to exchange of impurity
spin fluctuations will lead to%
\begin{equation}
\Sigma _{\sigma }(\omega ,T)\sim g^{2}\left( \omega \ln \frac{x}{D}-i\frac{%
\pi }{2}x\right) ,
\end{equation}%
where $x=\max (|\omega |,T)$ and $g$ is a coupling constant. This is a kind
of local marginal Fermi liquid behavior \cite{varma}. Furthermore, a
logarithmic correction to the linear-in-temperature specific heat can also
be obtained within the one-loop approximation.

\subsection{Effects of the external magnetic field}

In the presence of a magnetic field, an additional term $-hS^{z}$ lifts the
Ising degeneracy so the singularity at the Fermi level is suppressed. The
diagonal impurity retarded GFs are given by%
\begin{eqnarray*}
G_{1}(\omega ) &=&\frac{\left( \omega -\lambda +i\Gamma \right) \left(
\omega -\lambda -h+i\Gamma \right) +\Gamma ^{2}/2}{\left( \omega -\lambda
+i\Gamma \right) \left[ \left( \omega -\lambda \right) ^{2}+i2\Gamma \left(
\omega -\lambda \right) -h^{2}\right] }, \\
G_{0}(\omega ) &=&\frac{\left( \omega -\lambda +i\Gamma \right) ^{2}-h^{2}}{%
\left( \omega -\lambda +i\Gamma \right) \left[ \left( \omega -\lambda
\right) ^{2}+i2\Gamma \left( \omega -\lambda \right) -h^{2}\right] }, \\
G_{-1}(\omega ) &=&\frac{\left( \omega -\lambda +i\Gamma \right) \left(
\omega -\lambda +h+i\Gamma \right) +\Gamma ^{2}/2}{\left( \omega -\lambda
+i\Gamma \right) \left[ \left( \omega -\lambda \right) ^{2}+i2\Gamma \left(
\omega -\lambda \right) -h^{2}\right] }.
\end{eqnarray*}
If the magnetic field is weak, i.e., $h\ll \Gamma $, the impurity spectral
function exhibits three Lorentzian resonances at the same energy $\omega
=\lambda $, with different half widths $\Gamma $, and $\left( \Gamma \pm 
\sqrt{\Gamma ^{2}-h^{2}}\right) $, while the $\delta $-resonance disappears.
At $T=0$, $\lambda \approx 0$, and the impurity spin susceptibility is
derived as 
\begin{equation}
\chi _{d}\approx \frac{1}{\pi \Gamma }\left[ \ln \left( \frac{2\Gamma }{h}%
\right) -2\right] ,
\end{equation}
which displays a logarithmic dependence on $h$, indicating a local quantum
critical point at $T=0$ and $h=0$. When the magnetic field is strong enough,
i.e., $h>\Gamma $, the impurity spectral function shows three separated
Lorentzian resonances at $\omega =\lambda $ and $\omega =\lambda \pm \sqrt{%
h^{2}-\Gamma ^{2}}$, with the same half width $\Gamma $. For $h=\Gamma $,
three Lorentzian resonances with a half width $\Gamma $ reconcile again at $%
\omega =\lambda $.

\section{Effective model for spin-1/2 two Kondo impurity model}

The above described singular behavior in the underscreened spin-1 Kondo
impurity model only arises when $S>1/2$, while the real Kondo lattice
systems usually involve spin-1/2 magnetic impurities. However, it is also
possible for the underscreening to be an intrinsic feature around the
quantum critical point \cite{gegenwart}. The simplest model for such a study
is a spin-1/2 two magnetic Kondo impurity model \cite{jones,affleck},%
\begin{equation}
H=\sum_{\mathbf{k},\sigma }\epsilon _{\mathbf{k}}C_{\mathbf{k},\sigma
}^{\dagger }C_{\mathbf{k},\sigma }+J\left[ \mathbf{S}_{1}\cdot \mathbf{s}(%
\mathbf{r}_{1})+\mathbf{S}_{2}\cdot \mathbf{s}(\mathbf{r}_{2})\right] +I%
\mathbf{S}_{1}\cdot \mathbf{S}_{2},
\end{equation}%
where $\mathbf{S}_{1}$ and $\mathbf{S}_{2}$ are localized spin-1/2 moments
set at $\mathbf{r}_{1}=\mathbf{R}/2$ and $\mathbf{r}_{2}=-\mathbf{R}/2$,
respectively, the Kondo coupling is antiferromagnetic $J>0$, and a direct
inter-impurity interaction $I$ is included. When the inter-impurity
antiferromagnetic coupling $I/T_{k}$ is very large ($T_{k}$ is the single
impurity Kondo temperature), two spin-1/2 impurities lock into a singlet
state, and there is no Kondo effect and the conduction electrons are
completely unaffected by the impurities. Conversely, for a large and
ferromagnetic coupling $I/T_{k}$, the two impurities become an effective $S=1
$ magnetic impurity, and a fully screened Kondo behavior occurs at low
temperatures due to the presence of the two conduction electron channels.
However, the most intriguing behavior, as shown by the numerical RG studies,%
\cite{jones} appears in-between these two stable limiting phases, at a
plausible quantum critical point. Namely, the impurity singlet and triplet
states become \textit{degenerate} at a critical value of the inter-impurity
coupling $I_{c}\sim 2.2T_{K}$, displaying a local quantum critical behavior
with partially screened moment.

\subsection{Coupled representation for the spin-1/2 impurities}

Using the SU(2) pseudo-fermion representation for the spin-1/2 impurities,
the Kondo spin exchange interactions are written in the Coqblin-Schrieffer
form, 
\begin{eqnarray*}
H &=&\sum_{\mathbf{k},\sigma }\epsilon _{\mathbf{k}}C_{\mathbf{k},\sigma
}^{\dagger }C_{\mathbf{k},\sigma }+\frac{I}{2}\sum_{\sigma ,\sigma ^{\prime
}}d_{1,\sigma }^{\dagger }d_{1,\sigma ^{\prime }}d_{2,\sigma ^{\prime
}}^{\dagger }d_{2,\sigma } \\
&&-\frac{J}{2}\sum_{\sigma ,\sigma ^{\prime }}\left( d_{1,\sigma }^{\dagger
}d_{1,\sigma ^{\prime }}C_{\mathbf{r}_{1},\sigma }C_{\mathbf{r}_{1},\sigma
^{\prime }}^{\dagger }+d_{2,\sigma }^{\dagger }d_{2,\sigma ^{\prime }}C_{%
\mathbf{r}_{2},\sigma }C_{\mathbf{r}_{2},\sigma ^{\prime }}^{\dagger
}\right) ,
\end{eqnarray*}
where the potential scattering terms have been neglected. Moreover, two
constraints $\sum_{\sigma }d_{1,\sigma }^{\dagger }d_{1,\sigma }=1$ and $%
\sum_{\sigma }d_{2,\sigma }^{\dagger }d_{2,\sigma }=1$ have to be imposed.
In general, the two spin-1/2 operators can also be represented by four
coupled states 
\begin{eqnarray*}
&&|\uparrow ;\uparrow \rangle =f_{1}^{\dagger }|0\rangle ,
|\uparrow;\downarrow \rangle =f_{2}^{\dagger }|0\rangle , \\
&&|\downarrow ;\uparrow \rangle =f_{3}^{\dagger }|0\rangle , |\downarrow ;
\downarrow \rangle =f_{4}^{\dagger }|0\rangle ,
\end{eqnarray*}
which can be described by a local four-component fermion and the constraints
are replaced by $\sum_{\alpha }f_{\alpha }^{\dagger }f_{\alpha }=1$. Using
the projection procedure, the two impurity Kondo model can be expressed as 
\begin{widetext}
\begin{eqnarray}
H &=&\sum_{\mathbf{k},\sigma }\epsilon _{\mathbf{k}}C_{\mathbf{k},\sigma
}^{\dagger }C_{\mathbf{k},\sigma }-\frac{J}{2}\left[ \left( f_{1}^{\dagger
}C_{\mathbf{r}_{1},\uparrow }+f_{3}^{\dagger }C_{\mathbf{r}_{1},\downarrow
}\right) \left( C_{\mathbf{r}_{1},\uparrow }^{\dagger }f_{1}+C_{\mathbf{r}%
_{1},\downarrow }^{\dagger }f_{3}\right) +\left( f_{2}^{\dagger }C_{\mathbf{r%
}_{1},\uparrow }+f_{4}^{\dagger }C_{\mathbf{r}_{1},\downarrow }\right)
\left( C_{\mathbf{r}_{1},\uparrow }^{\dagger }f_{2}+C_{\mathbf{r}%
_{1},\downarrow }^{\dagger }f_{4}\right) \right.  \nonumber\\
&&+\left. \left( f_{1}^{\dagger }C_{\mathbf{r}_{2},\uparrow
}+f_{2}^{\dagger }C_{\mathbf{r}_{2},\downarrow }\right) \left(
C_{\mathbf{r}_{2},\uparrow}^{\dagger}f_{1}+C_{\mathbf{r}_{2},\downarrow
}^{\dagger }f_{2}\right)
+\left( f_{3}^{\dagger }C_{\mathbf{r}_{2},\uparrow }+f_{4}^{\dagger }C_{%
\mathbf{r}_{2},\downarrow }\right) \left( C_{\mathbf{r}_{2},\uparrow
}^{\dagger }f_{3}+C_{\mathbf{r}_{2},\downarrow }^{\dagger }f_{4}\right) %
\right] \nonumber \\
&&+\frac{I}{2}\left( f_{1}^{\dagger }f_{1}+f_{4}^{\dagger
}f_{4}+f_{2}^{\dagger }f_{3}+f_{3}^{\dagger }f_{2}\right),
\end{eqnarray}
where the direct inter-impurity coupling has been transformed into a
quadratic form, and can be considered exactly.

\subsection{Effective resonant level model}

In order to derive an effective model to describe the local quantum critical
point, we employ a similar method as in the functional integral approach for
the spin-1/2 Kondo impurity model \cite{read}, and a saddle point solution
can be deduced by introducing an effective hybridization amplitude: 
\[
\langle f_{1}^{\dagger }C_{\mathbf{r}_{1},\uparrow }+f_{3}^{\dagger }C_{%
\mathbf{r}_{1},\downarrow }\rangle =\langle f_{2}^{\dagger }C_{\mathbf{r}%
_{1},\uparrow }+f_{4}^{\dagger }C_{\mathbf{r}_{1},\downarrow }\rangle
=\langle f_{1}^{\dagger }C_{\mathbf{r}_{2},\uparrow }+f_{2}^{\dagger }C_{%
\mathbf{r}_{2},\downarrow }\rangle =\langle f_{3}^{\dagger }C_{\mathbf{r}%
_{2},\uparrow }+f_{4}^{\dagger }C_{\mathbf{r}_{2},\downarrow }\rangle =-V, 
\]%
where the spin rotational symmetry and parity symmetry between the magnetic
impurities ($\mathbf{S}_{1}\Leftrightarrow \mathbf{S}_{2}$)\ have been
assumed. Using linear combinations of the impurity spin anti-parallel
states, $\widetilde{f_{2}}=(f_{2}+f_{3})/\sqrt{2}$ \ and $\widetilde{f}%
_{3}=(f_{2}-f_{3})/\sqrt{2}$, an effective resonant level model can be
derived: 
\begin{eqnarray}
&&H_{eff}=\sum_{\mathbf{k},\sigma }\epsilon _{\mathbf{k}}C_{\mathbf{k}%
,\sigma }^{\dagger }C_{\mathbf{k},\sigma }+\frac{JV}{\sqrt{2N}}\sum_{\mathbf{%
k,\sigma }}\left( C_{\mathbf{k},\sigma }^{\dagger }\widetilde{f}_{2}\cos 
\frac{\mathbf{k\cdot R}}{2}+h.c.\right) +\frac{JV}{\sqrt{N}}\sum_{\mathbf{k}}%
\left[ \left( C_{\mathbf{k},\uparrow }^{\dagger }f_{1}+C_{\mathbf{k}%
,\downarrow }^{\dagger }f_{4}\right) \cos \frac{\mathbf{k\cdot R}}{2}+h.c.%
\right]  \nonumber \\
&&-\frac{JV}{\sqrt{2N}}\sum_{\mathbf{k,\sigma }}\left( i\sigma C_{\mathbf{k}%
,\sigma }^{\dagger }\widetilde{f}_{3}\sin \frac{\mathbf{k\cdot R}}{2}%
+h.c.\right) +\left( \lambda -I/2\right) \widetilde{f}_{3}^{\dagger }%
\widetilde{f}_{3}+\left( \lambda +I/2\right) \left( f_{1}^{\dagger
}f_{1}+f_{4}^{\dagger }f_{4}+\widetilde{f}_{2}^{\dagger }\widetilde{f}%
_{2}\right) +(2JV^{2}-\lambda ),
\end{eqnarray}%
%
where the local constraint has been implemented by a Lagrange multiplier $%
\lambda $. Then the conduction electrons in three spatial dimension can be
reduced and divided into symmetric and antisymmetric channels, defined by 
\begin{equation}
C_{k,+,\sigma }=\frac{1}{\sqrt{N_{+}(k)}}\int \frac{d\Omega _{\mathbf{k}}}{%
4\pi }C_{\mathbf{k},\sigma }\cos \frac{\mathbf{k\cdot R}}{2},\text{ \ }%
C_{k,-,\sigma }=\frac{i}{\sqrt{N_{-}(k)}}\int \frac{d\Omega _{\mathbf{k}}}{%
4\pi }C_{\mathbf{k},\sigma }\sin \frac{\mathbf{k\cdot R}}{2},
\end{equation}%
where $N_{\pm }(k)=\frac{1}{2}\left( 1\pm {\sin kR}/(kR)\right) $ and the
effective resonant level model becomes a one-dimensional system with two
separated conduction electron channels interacting with two magnetic
impurities $H_{eff}=H_{+}+H_{-}+(2JV^{2}-\lambda )$, where 
\begin{eqnarray}
H_{+} &=&\sum_{k}\epsilon _{k}C_{k,+,\sigma }^{\dagger }C_{k,+,\sigma }+%
\frac{JV}{\sqrt{N}}\sum_{k}\sqrt{N_{+}(k)}\left[ \left( C_{k,+,\uparrow
}^{\dagger }f_{1}+C_{k,+,\downarrow }^{\dagger }f_{4}\right) +h.c.\right] 
\nonumber \\
&&+\frac{JV}{\sqrt{2N}}\sum_{k,\mathbf{\sigma }}\sqrt{N_{+}(k)}\left(
C_{k,+,\sigma }^{\dagger }\widetilde{f}_{2}+h.c.\right) +\left( \lambda
+I/2\right) \left( f_{1}^{\dagger }f_{1}+f_{4}^{\dagger }f_{4}+\widetilde{f}%
_{2}^{\dagger }\widetilde{f}_{2}\right) ,  \nonumber \\
H_{-} &=&\sum_{k}\epsilon _{k}C_{k,-,\sigma }^{\dagger }C_{k,-,\sigma }-%
\frac{JV}{\sqrt{2N}}\sum_{k\mathbf{,\sigma }}\sqrt{N_{-}(k)}\left( \sigma
C_{k,-,\sigma }^{\dagger }\widetilde{f}_{3}+h.c.\right) +\left( \lambda
-I/2\right) \widetilde{f}_{3}^{\dagger }\widetilde{f}_{3}.
\end{eqnarray}%
\end{widetext} It is clearly seen that $\widetilde{f}_{3}^{\dagger
}|0\rangle $ describes the impurity singlet state with energy $\left(
\lambda -I/2\right) $, and hybridizes with the \textit{antisymmetric}
channel of the conduction electrons, while $f_{1}^{\dagger }|0>$, $%
f_{4}^{\dagger }|0>$, and $\widetilde{f}_{2}^{\dagger }|0>$ describe the
impurity triplet states with energy $\left( \lambda +I/2\right) =\epsilon
_{f}$ and couple to the \textit{symmetric} channel of the conduction
electrons. The effective resonant level model in the symmetric channel
shares a \textit{similar} form of the effective model as for the isotropic
underscreened spin-1 Kondo impurity model, except for the momentum
dependence of the hybridization integrals.

\subsection{Impurity spectral function}

Now the effective Hamiltonian can be diagonalized as before, and the
retarded impurity quasiparticle GFs are obtained: 
\begin{eqnarray}
G_{1}(\omega )&=&G_{4}(\omega )=\frac{1/4}{\omega -\epsilon _{f}}+\frac{1/4}{%
\omega -\epsilon _{f}-2\Delta _{+}+2i\Gamma _{+}}  \nonumber \\
&&+\frac{1/2}{\omega -\epsilon _{f}-\Delta _{+}+i\Gamma _{+}},  \nonumber \\
G_{2}(\omega ) &=&\frac{1/2}{\omega -\epsilon _{f}}+\frac{1/2}{\omega
-\epsilon _{f}-2\Delta _{+}+2i\Gamma _{+}},  \nonumber \\
G_{3}(\omega ) &=&\frac{1}{\omega -\epsilon _{f}+I-\Delta _{-}+i\Gamma _{-}},%
\text{ }
\end{eqnarray}
where the self-energy corrections are involved 
\begin{eqnarray}
&& \Delta _{\pm }(\omega )=\frac{\left( JV\right) ^{2}}{N}\sum_{k}\frac{%
N_{\pm }(k)}{\omega -\epsilon _{k}},  \nonumber \\
&&\Gamma _{\pm }=\pi \rho (k_{f})N_{\pm }(k_{f})(JV)^{2}.
\end{eqnarray}
Thus, the impurity quasiparticle spectral function is given by the imaginary
parts of the retarded GFs, i.e. \begin{widetext} 
\begin{equation}
A_{\mathrm{imp}}(\omega ) =\delta \left(\omega-\epsilon_{f}\right) + \frac{%
\Gamma _{+}/\pi }{\left( \omega -\epsilon _{f}-\Delta _{+}\right)
^{2}+\Gamma _{+}^{2}}+\frac{2\Gamma _{+}/\pi }{\left( \omega -\epsilon
_{f}-2\Delta _{+}\right) ^{2}+4\Gamma _{+}^{2}}+\frac{\Gamma _{-}/\pi }{%
\left( \omega -\epsilon _{f}+I-\Delta _{-}\right) ^{2}+\Gamma _{-}^{2}},
\end{equation}
\end{widetext} where four resonances appear in the impurity spectral
function, and the $\delta $-resonance persists at the impurity energy level $%
\omega =\epsilon _{f}$, leading to a singular behavior for the low energy
quasiparticles.

When the inter-impurity coupling is antiferromagnetic $I>0$, the impurity
singlet state is the lowest energy state, while for the ferromagnetic
inter-impurity interaction $I<0$, the impurity triplet state has the lowest
energy. In general, the energy levels of both the impurity singlet and
triplet states should be \textit{strongly} renormalized by both the
inter-impurity interaction and the couplings with the conduction electrons.
Such a renormalization effects on the impurity energy levels are not fully
considered in the present treatments. However, in the symmetric channel of
the effective model, we have found a $\delta $-resonance singularity
corresponding to the partially screened impurity triplet state, which is
similar to what we find in the underscreened spin-1 Kondo impurity model.
Therefore, we believe that the singular behavior of the spin-1/2 two Kondo
impurity model around the nontrivial quantum critical point is closely
related to the low-energy physics of the spin-1 underscreened Kondo model.
This indicates that the peculiar behavior observed in a large class of heavy
fermion metals\cite{schroder,gegenwart} might be related to the physics of
underscreened Kondo model.

\section{Quantum dots with even number of electrons}

To experimentally observe the singular behavior inherent to the
underscreened spin-1 Kondo impurity model, the mesoscopic quantum dot
systems with even number of electrons are among the promising candidates %
\cite{hofstetter,coleman-2004}. In the presence of a large Hund's rule
coupling between two topmost dot electrons, an effective two coupled
spin-1/2 Kondo impurity model has been derived by Kikoin and Avishai \cite%
{ka}. 
\begin{equation}
H=\sum_{\mathbf{k},\sigma }\epsilon _{\mathbf{k}}C_{\mathbf{k},\sigma
}^{\dagger }C_{\mathbf{k},\sigma }+\left( J_{1}\mathbf{S}_{1}+J_{2}\mathbf{S}%
_{2}\right) \cdot \mathbf{s}(\mathbf{0})-I\mathbf{S}_{1}\cdot \mathbf{S}_{2},
\end{equation}%
where $J_{1}$ and $J_{2}$ are two positive antiferromagnetic Kondo
couplings, the Hund's coupling $I$ is ferromagnetic, and the two magnetic
moments are sitting on the same site. Using the SU(2) pseudo-fermion
representation for the spin-1/2 operators, the Kondo spin exchanges are
written in the Coqblin-Schrieffer form, 
\begin{eqnarray}
H &=&\sum_{\mathbf{k},\sigma }\epsilon _{\mathbf{k}}C_{\mathbf{k},\sigma
}^{\dagger }C_{\mathbf{k},\sigma }-\frac{I}{2}\sum_{\sigma ,\sigma ^{\prime
}}d_{1,\sigma }^{\dagger }d_{1,\sigma ^{\prime }}d_{2,\sigma ^{\prime
}}^{\dagger }d_{2,\sigma }  \nonumber \\
&&-\frac{1}{2}\sum_{\sigma ,\sigma ^{\prime }}\left( J_{1}d_{1,\sigma
}^{\dagger }d_{1,\sigma ^{\prime }}+J_{2}d_{2,\sigma }^{\dagger }d_{2,\sigma
^{\prime }}\right) C_{\sigma }C_{\sigma ^{\prime }}^{\dagger }
\end{eqnarray}%
where the potential scattering terms have been neglected. Two single
occupied constraints $\sum_{\sigma }d_{1,\sigma }^{\dagger }d_{1,\sigma }=1$
and $\sum_{\sigma }d_{2,\sigma }^{\dagger }d_{2,\sigma }=1$ are imposed. The
two spin-1/2 operators can also be represented by four states $|\uparrow
;\uparrow \rangle =f_{1}^{\dagger }|0\rangle $, $|\uparrow ;\downarrow
\rangle =f_{2}^{\dagger }|0\rangle $, $|\downarrow ;\uparrow \rangle
=f_{3}^{\dagger }|0\rangle $, $|\downarrow ;\downarrow \rangle
=f_{4}^{\dagger }|0\rangle $ and the constraints are replaced by $%
\sum_{\alpha }f_{\alpha }^{\dagger }f_{\alpha }=1$. Then the model can be
expressed as \begin{widetext} 
\begin{eqnarray}
H &=&\sum_{\mathbf{k},\sigma }\epsilon _{\mathbf{k}}C_{\mathbf{k},\sigma
}^{\dagger }C_{\mathbf{k},\sigma }-\frac{J_{1}}{2}\left[ \left(
f_{1}^{\dagger }C_{\uparrow }+f_{3}^{\dagger }C_{\downarrow }\right) \left(
C_{\uparrow }^{\dagger }f_{1}+C_{\downarrow }^{\dagger }f_{3}\right) +\left(
f_{2}^{\dagger }C_{\uparrow }+f_{4}^{\dagger }C_{\downarrow }\right) \left(
C_{\uparrow }^{\dagger }f_{2}+C_{\downarrow }^{\dagger }f_{4}\right) \right]
\\
&&-\frac{J_{2}}{2}\left[ \left( f_{1}^{\dagger }C_{\uparrow }+f_{2}^{\dagger
}C_{\downarrow }\right) \left( C_{\uparrow }^{\dagger }f_{1}+C_{\downarrow
}^{\dagger }f_{2}\right) +\left( f_{3}^{\dagger }C_{\uparrow
}+f_{4}^{\dagger }C_{\downarrow }\right) \left( C_{\uparrow }^{\dagger
}f_{3}+C_{\downarrow }^{\dagger }f_{4}\right) \right] -\frac{I}{2}\left(
f_{1}^{\dagger }f_{1}+f_{4}^{\dagger }f_{4}+f_{2}^{\dagger
}f_{3}+f_{3}^{\dagger }f_{2}\right) ,  \nonumber
\end{eqnarray}%
\end{widetext} Actually, using an SU(4) pseudo-fermion representation, both
spin-1/2 operators $\mathbf{S}_{1}$ and $\mathbf{S}_{2}$ are expressed
jointly \cite{gmzhang}, and the same form of the model Hamiltonian can be
obtained.

\subsection{Effective model Hamiltonian}

Implementing the constraint by a Lagrangian multiplier $\lambda $ and
introducing two effective hybridization variables: $\langle f_{1}^{\dagger
}C_{\uparrow }+f_{3}^{\dagger }C_{\downarrow }\rangle =\langle
f_{2}^{\dagger }C_{\uparrow }+f_{4}^{\dagger }C_{\downarrow }\rangle =-v_{1}$
and $\langle f_{1}^{\dagger }C_{\uparrow }+f_{2}^{\dagger }C_{\downarrow
}\rangle =\langle f_{3}^{\dagger }C_{\uparrow }+f_{4}^{\dagger
}C_{\downarrow }\rangle =-v_{2}$, the strong coupling MF approximation can
be carried out as before. With linear combinations of the impurity spin
anti-parallel states, $\widetilde{f_{2}}=(f_{2}+f_{3})/\sqrt{2}$ \ and $%
\widetilde{f}_{3}=(f_{2}-f_{3})/\sqrt{2}$, a modified effective resonant
level model can be derived: \begin{widetext} 
\begin{eqnarray}
H_{eff} &=&\sum_{\mathbf{k},\sigma }\epsilon _{\mathbf{k}}C_{\mathbf{k}%
,\sigma }^{\dagger }C_{\mathbf{k},\sigma }+\frac{V}{\sqrt{2N}}\sum_{\mathbf{%
k,\sigma }}\left( C_{\mathbf{k},\sigma }^{\dagger }\widetilde{f}%
_{2}+h.c.\right) +\frac{V}{\sqrt{N}}\sum_{\mathbf{k}}\left( C_{\mathbf{k}%
,\uparrow }^{\dagger }f_{1}+C_{\mathbf{k},\downarrow }^{\dagger
}f_{4}+h.c.\right)  \nonumber \\
&&+\frac{Vr}{\sqrt{2N}}\sum_{\mathbf{k,\sigma }}\left( \sigma C_{\mathbf{k}%
,\sigma }^{\dagger }\widetilde{f}_{3}+h.c.\right) +\left( \epsilon
_{f}+I\right) \widetilde{f}_{3}^{\dagger }\widetilde{f}_{3}+\epsilon
_{f}\left( f_{1}^{\dagger }f_{1}+f_{4}^{\dagger }f_{4}+\widetilde{f}%
_{2}^{\dagger }\widetilde{f}_{2}\right)
+(J_{1}v_{1}^{2}+J_{2}v_{2}^{2}-\lambda ),
\end{eqnarray}%
\end{widetext} where $V=(J_{1}v_{1}+J_{2}v_{2})/2$, $%
r=(J_{1}v_{1}-J_{2}v_{2})/(J_{1}v_{1}+J_{2}v_{2})$, $\epsilon _{f}=\lambda
-I/2$, and only one conduction electron channel effectively couples to the
dot electrons. It is clearly seen that $\widetilde{f}_{3}^{\dagger
}|0\rangle $ describes the impurity singlet state with energy $\left(
\epsilon _{f}+I\right) $, and hybridizes with the spin-up and -down
conduction electrons, while the impurity triplet state with energy $\epsilon
_{f}$ couples to the conduction electrons as discussed above. Since the
inter-impurity ferromagnetic coupling $I$ is large enough, the impurity
singlet energy level lies far above $\epsilon _{f}$ while the triplet state
lies at $\epsilon _{f}$. The effective resonant level model in the limit of $%
J_{1}=J_{2}$ reduces to the effective model for the isotropic underscreened
spin-1 Kondo impurity model in the strong coupling limit, where the impurity
singlet state completely decouples from the conduction electrons.

\subsection{Dot electron spectral function and spin susceptibilities}

Now the effective Hamiltonian is diagonalized, and the dot electron spectral
function is thus given by \begin{widetext} 
\begin{equation}
A_{\mathrm{dot}}(\omega )=\delta \left( \omega -\epsilon _{f}\right) +\frac{1%
}{\pi }\frac{2\Gamma }{\left( \omega -\epsilon _{f}\right) ^{2}+4\Gamma ^{2}}%
+\frac{1}{\pi }\frac{\Gamma \left[ (1+r^{2})\left( \omega -\epsilon
_{f}\right) ^{2}-2I\left( \omega -\epsilon _{f}\right) +I^{2}\right] }{%
\left( \omega -\epsilon _{f}\right) ^{2}\left( \omega -\epsilon
_{f}-I\right) ^{2}+\Gamma ^{2}\left[ (1+r^{2})\left( \omega -\epsilon
_{f}\right) -I\right] ^{2}},
\end{equation}%
\end{widetext} where four resonances appear in the dot spectral function,
and the $\delta $-resonance persists at $\omega =\epsilon _{f}$, leading to
a singular behavior for the dot electrons the same way as in the
underscreened spin-1 Kondo impurity model. Since the uniform spin and
staggered spin density operators are expressed as 
\begin{eqnarray}
\left( S_{1}^{z}+S_{2}^{z}\right) &=&f_{1}^{\dagger }f_{1}-f_{4}^{\dagger
}f_{4},  \nonumber \\
\left( S_{1}^{z}-S_{2}^{z}\right) &=&\widetilde{f}_{2}^{\dagger }\widetilde{f%
}_{3}+\widetilde{f}_{3}^{\dagger }\widetilde{f}_{2},
\end{eqnarray}
the corresponding spin susceptibilities can be evaluated and found: $\mathtt{%
Im}\chi _{\mathrm{imp}}^{u}(\omega ,T)$, $\mathtt{Im}\chi _{\mathrm{imp}%
}^{s}(\omega ,T)\propto \tanh \left( {\omega }/(2k_{B}T)\right) $, from
which the corresponding real part dynamic susceptibilities display a
logarithmic dependence on $\max (\omega ,T)$ as for the spin-1 underscreened
Kondo impurity model. We can also show that the singularity of the $\delta $%
-resonance is only associated with the triplet state. Thus, we can interpret
the large zero-bias anomalies of the differential conductance experimentally
observed on the quantum dot with even occupied electrons without a magnetic
field \cite{schmid-kogan} as a manifestation of the residual Z$_{2}$ Ising
degeneracy of the bound state composed of the triplet dot electron state
with the conduction electrons in the leads.

\section{Conclusion}

To conclude, by proposing an effective resonant level model in the strong
coupling limit for the spin-1 underscreened Kondo impurity model, we have
shown that a local quantum critical behavior (logarithmic divergence of the
local dynamic spin susceptibility) is induced by the formation of a bound
state with a partially screened magnetic moment. We believe this is a 
\textit{generic} feature of all underscreened Kondo models. The \textit{%
quantum} nature of the remanent magnetic moment is explicitly exhibited
through the symmetry determined degeneracy. A direct experimental
observation of this quantum critical behavior might be possible in the
quantum dot systems with even occupied electrons, and its implications
related to some peculiar properties of f-electron heavy fermion systems are
also indicated through the exploration of the spin-1/2 two Kondo impurity
model.

\begin{acknowledgments}
G.-M. Zhang is grateful to Professor H. Shiba for his encouragement during
this work and acknowledges the support of NSF-China (Grant No.10125418) and
the Special Fund for Major State Basic Research Projects of China (Grant
No.G200067107). L. Yu also acknowledges the support of NSF-China.
\end{acknowledgments}

\end{document}